**Demonstration of a uniform, high-pressure, high-temperature gas cell with a dual frequency comb absorption spectrometer**

Ryan K. Cole*, Anthony D. Draper, Paul J. Schroeder, Cameron M. Casby, Amanda S. Makowiecki, Sean C. Coburn, Julie E. Steinbrenner, Nazanin Hoghooghi, Gregory B. Rieker*

*Precision Laser Diagnostics Laboratory, Department of Mechanical Engineering*
*University of Colorado Boulder, Boulder, CO 80309 USA*
*ryan.cole@colorado.edu, greg.rieker@colorado.edu

**Abstract**

Accurate absorption models for gases at high pressure and temperature support advanced optical combustion diagnostics and aid in the study of harsh planetary atmospheres. Developing and validating absorption models for these applications requires recreating the extreme temperature and pressure conditions of these environments in static, uniform, well-known conditions in the laboratory. Here, we present the design of a new gas cell to enable reference-quality absorption spectroscopy at high pressure and temperature. The design centers on a carefully controlled quartz sample cell housed at the core of a pressurized ceramic furnace. The half-meter sample cell is relatively long compared to past high-pressure and -temperature absorption cells, and is surrounded by a molybdenum heat spreader that enables high temperature uniformity over the full length of the absorbing gas. We measure the temperature distribution of the sample gas using *in situ* thermocouples, and fully characterize the temperature uniformity across a full matrix of temperatures and pressures up to 1000 K and 50 bar. The results demonstrate that the new design enables highly-uniform and precisely-known temperature and pressure conditions across the full absorbing path length. Uniquely, we test the new gas cell with a broadband (~2500 cm$^{-1}$), high-resolution (0.0066 cm$^{-1}$) dual frequency comb spectrometer that enables highly resolved absorption spectroscopy across a wide range of temperature and pressure conditions. With this carefully characterized system, we measure the spectrum of $CO_2$ between 6800 and 7000 cm$^{-1}$ at pressures between 0.2 and 20 bar, and temperatures up to 1000 K. The measurements reveal discrepancies from spectra predicted by the HITRAN2016 database with a Voigt line shape at both low- and high-pressure conditions. These results motivate future work to expand absorption models and databases to accurately model high-pressure and -temperature spectra in combustion and planetary science research.

**Keywords:** Optical gas cell, absorption spectroscopy, high temperature, high pressure, dual frequency comb spectroscopy

## 1. Introduction

Absorption spectroscopy is an important diagnostic technique in fields ranging from planetary science to combustion. In absorption spectroscopy, the thermodynamic properties of a gas (e.g. an exoplanetary atmosphere or inside of an internal combustion engine) can be determined from the intensity and shape of a measured absorption spectrum using absorption models and tabulated spectral databases. Both planetary science research and combustion diagnostics require accurate absorption models for gases under high-pressure and -temperature conditions. In planetary science, absorption models are used to interpret observations of emission and absorption spectra of exotic atmospheres and as inputs to radiative transfer models [1,2]. In combustion diagnostics, condition-specific absorption models are fit to data from optical sensors to measure gas properties in high-pressure and -temperature systems [3]. In both applications, inaccuracies in the absorption model or database limit the interpretation of physical and chemical information from spectra measured in high-pressure and -temperature systems.

High temperatures and pressures pose significant challenges for the development of accurate absorption models and databases. At high temperatures, absorption databases must be updated with temperature-dependent and species-specific line shape parameters (e.g. pressure broadening and shift coefficients) as well as positions and intensities of transitions from quantum states that become populated at high temperatures. High pressures complicate absorption models due to the increased importance of non-Lorentzian collisional effects such as line mixing, collision-induced absorption, and the breakdown of the impact approximation. These effects, which themselves are temperature dependent, can cause significant errors when omitted from absorption models at high pressures [4,5]. A number of databases and large-scale calculations have been compiled to expand available data at high temperature [e.g. 6–11], and numerous other studies have been devoted to modeling collisional effects at high pressures (recently summarized in [4,12]). However, only a fraction of this past work considers simultaneously high-pressure and -temperature conditions. As such, further laboratory studies are needed in these conditions to expand available data, validate calculations, and continue to advance absorption models for high temperatures and pressures.

A significant challenge facing laboratory spectroscopy at high pressure and temperature is the difficulty of creating optically accessible gas cells capable of generating uniform, precisely known, high-pressure and -temperature conditions in the laboratory. Accurate spectroscopy in these conditions places demanding requirements on the gas cell design. For studies of collisional effects such as line mixing or collision induced absorption, these gas cells must enable optical access at pressures where these effects become pronounced in measured spectra. Depending on the collisional physics and the molecule in question, these pressures can range anywhere from several to 100+ bar. Further, to study the temperature dependence of the collisional effects, the gas cell must be capable of accessing very high temperatures while at high-pressure conditions. Critically, high-temperature conditions must be generated in such a way that temperature gradients in the absorbing gas are minimized. Significant temperature variations lead to errors in the interpretation of measured spectra because the absorbing molecules do not occupy the quantum states predicted by a single temperature (breaking a key assumption in most fitting and modeling routines). These requirements are further complicated by the fact that many reference absorption measurements benefit from relatively long path lengths (tens of centimeters or more) in order to measure weak absorption by high-temperature gases and in order to accommodate highly dilute mixtures of absorbing gases that are ideal for model development [4,5]. As such, gas cells for accurate laboratory spectroscopy at high pressure and temperature must be designed to achieve the widest possible range in accessible pressures and temperatures while also balancing the need to maximize the absorbing path length and minimize temperature variations and uncertainty in the sample gas conditions.

Following these design constraints, a number of gas cells for absorption spectroscopy at high pressure and temperature have been designed and evaluated, as discussed in detail in Section 2 [13–21]. Generally, most recent cell designs fall into two categories: designs that directly heat the high-pressure vessel, and 'cell-in-cell' designs that place the heating elements and sample cell inside of the high-pressure vessel. The first design has been shown to be capable of sustaining very high-pressure and -temperature conditions (1000+ K and 100+ bar [19,21]), but requires that the pressure cell windows maintain a seal at high temperature, or that the windows are very thick (15+ cm [20,21]) so that they extend into the cold region where a low-temperature seal can be implemented. The cell-in-cell design, on the other hand, can offer high uniformity and thin windows by using separate, cold, high-pressure windows outside of the sample cell. However, this design has so far only been demonstrated at modest pressures and temperatures (up to 565 K and 20 bar [18]).

In this paper, we present the design and validation of a cell-in-cell design for laboratory spectroscopy in conditions exceeding 1000 K and 50 bar with high relative temperature uniformity and a long absorbing path length. We test the gas cell design and fully characterize the temperature uniformity along the 45.82 cm absorbing path length across a full matrix of temperatures and pressures up to 1000 K and 50 bar. We combine the gas cell with a broadband (~2500 cm$^{-1}$), high-resolution (0.0066 cm$^{-1}$) dual frequency comb spectrometer. This spectrometer combines the spectral resolution required to resolve narrow absorption features at low pressures with the bandwidth needed to measure entire absorption bands that have been severely broadened and blended at high pressure and temperature. With this combination, we measure the broadband, high-resolution spectrum of $CO_2$ between 6800 cm$^{-1}$ and 7000 cm$^{-1}$ across a wide range of temperature and pressure conditions up to 1000 K and 15 bar.

## 2. Existing gas cells for absorption spectroscopy at high temperature and pressure

A number of gas cell designs have been demonstrated across a range of high-pressure and -temperature conditions [13–21]. As many of these designs were recently summarized in Refs. [20,21], in this section we will only give a brief review of the most recent gas cell designs.

Recently, Refs. [19–21] published gas cells with designs based on directly heating the high-pressure vessel. The design employs a high-temperature-capable pressure cell (e.g. Inconel-based) that is housed inside a high-temperature heating element. The edges of the pressure cell extend beyond the heated area and are water cooled in order to maintain a safe and effective pressure seal. In order to mitigate absorption in these edge regions with the highest thermal gradient, the absorbing region is confined to the center of the pressure cell using long windows that extend into the more uniform, high-temperature core. For example, in Ref. [19], thick sapphire windows bonded to high-pressure-capable ceramic tubes are inserted into the pressure vessel and sealed to the end flanges where they are left open to non-absorbing atmospheric gas. Similarly, the design of Refs. [20,21] uses lengthy rods of transmissive glass bonded to the end flanges of the pressure vessel to restrict the absorbing sample to the more uniform region at the core of the pressure vessel. In both Refs. [20,21], $CaF_2$ is used as the material for the glass rods, which optimizes the gas cell for mid-infrared spectroscopy up to 8 μm. Notably, both designs require the glass elements affording optical access into the pressurized region to be subject to both high temperatures and a large pressure drop. Spectroscopy using this cell design has been demonstrated in conditions in excess of 1000 K and 100 bar [19,21].

The second gas cell design approach (similar to the design presented in this paper) has been described in Refs. [18,22,23]. In this approach, the high-temperature heating elements are housed inside of the pressurized volume. The absorbing gas is confined in the core of the high-temperature, high-pressure region in a transparent cell. The pressures inside and outside of the glass sample cell are maintained on separate gas lines so that the sample cell experiences only a slight pressure differential and thus does not rupture even at very high pressures. The bath gas surrounding the sample cell is non-absorbing so that absorption only occurs in the gas in the sample cell. The pressure vessel is sealed by a pair of windows outside of the high-temperature region so that these windows experience only a pressure drop and no significant temperature load. This approach can use much thinner windows (which is important as some window materials, such as sapphire, become absorptive at high temperature), and allows for good isolation of the sample cell in the high-temperature core. These benefits come at the cost of increased complexity in the overall system. This cell-in-cell design has been applied to path lengths (length of the sample cell) of 2 cm, and has been used to record spectra in conditions up to 565 K and 20 bar [18].

## 3. Experimental Design

The gas cell we present here applies the principles of the cell-in-cell design discussed in the preceding section, and has been optimized to achieve high temperature uniformity over a path length significantly longer than previous gas cells. Figure 1 shows an overview of the gas cell design and experiment including critical dimensions.

The gas cell consists of a custom steel pressure vessel that is designed to withstand pressures of 100 bar at skin temperatures up to 300 °C. Heating is provided by a single-zone ceramic furnace that is housed inside of the pressure vessel. The ceramic furnace uses nine silicon carbide heating rods with a total power of 16.9 kW to generate temperatures in excess of 1000 K across the 76 cm heated length of the furnace. The furnace is capped on both ends by two 7.6 cm thick discs of ceramic insulation, each with a ~4 cm hole at the center to allow optical access into the core of the furnace. The 45.82±0.09 cm fused quartz sample cell (4.6 cm diameter) is supported by ceramic rods on the center axis of the furnace, and defines the absorbing path length of the gas cell. Figure 2(a) shows a cross section of the ceramic furnace and sample cell. The sample cell has two quartz feed tubes at either end of the cell to enable flow-through operation. The gas lines extend out of the furnace and are connected to the gas handling system using a glass-to-metal transition. Optical access into the pressure vessel is enabled through a pair of sapphire windows with a nominal thickness of 1.1 cm. The windows are mounted to the outer flanges of the pressure vessel in commercial NPT housings (Rayotek Sight Windows) with a clear aperture of 6.6 cm. Both the sapphire windows and the quartz windows on the sample cell are manufactured with a slight wedge to mitigate etalon interference effects in measured spectra.

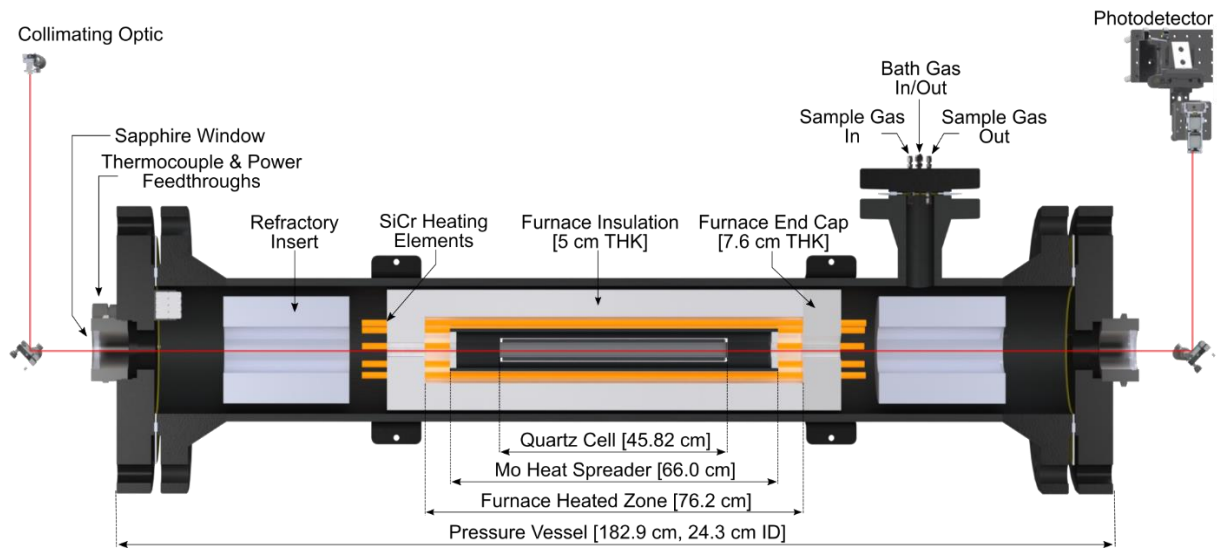

**Figure 1:** Schematic of the high-pressure and -temperature gas cell showing the primary components and critical dimensions. The gas cell consists primarily of a ceramic furnace housed inside of a custom pressure vessel. A 46-cm glass sample cell is placed at the core of the pressurized furnace, and surrounded by a molybdenum heat spreader to increase the temperature uniformity within the sample gas. Light from the spectrometer is passed axially though the gas cell to record the spectrum of the sample gas at the conditions at the core of the pressurized furnace.

The combination of the long path length with the need for high temperature uniformity poses a significant challenge. In order to optimize the temperature uniformity along the sample cell, the cell is surrounded in the furnace by a molybdenum tube (see Fig. 2(a)). The molybdenum tube acts as a heat spreader, providing an efficient pathway for the conduction of heat from the silicon carbide heating rods along the length of the sample cell. Molybdenum was chosen as the material for the heat spreader due to its unique combination of high working temperature and high thermal conductivity (~1/3 that of copper) [24]. In tests of the design, the addition of the heat spreader significantly increased the temperature uniformity when compared to trials where the heat spreader was not installed.

The temperature distribution of the sample cell is monitored by seven K-type thermocouples inside the ceramic furnace, shown in Figure 2(b). Five of these thermocouples are bolted to the interior of the molybdenum heat spreader to monitor its temperature distribution. The thermocouple at the center of the heat spreader is used to control the furnace set point. Two additional fine-wire thermocouples are fed into the sample cell through the gas lines to provide continuous, *in situ* monitoring of the edge temperatures of the sample gas during experiments.

The sample cell pressure is monitored by three pressure transducers optimized for different pressure ranges. Atmospheric and sub-atmospheric pressures are measured using a 0 – 1000 torr capacitance manometer (MKS Baratron 626C) with an accuracy of 0.25% of reading. Intermediate pressures (0 – 13.3 bar) are monitored using a capacitance manometer (MKS Baratron 121A) with an accuracy of 0.5% of reading. Finally, high pressures (0 – 100 bar) are measured using the third capacitance manometer (MKS Baratron 750C) accurate to 0.1% of full scale. A fourth pressure transducer (TE Connectivity, $\pm 0.2$ bar accuracy) monitors the bath gas pressure outside of the sample cell.

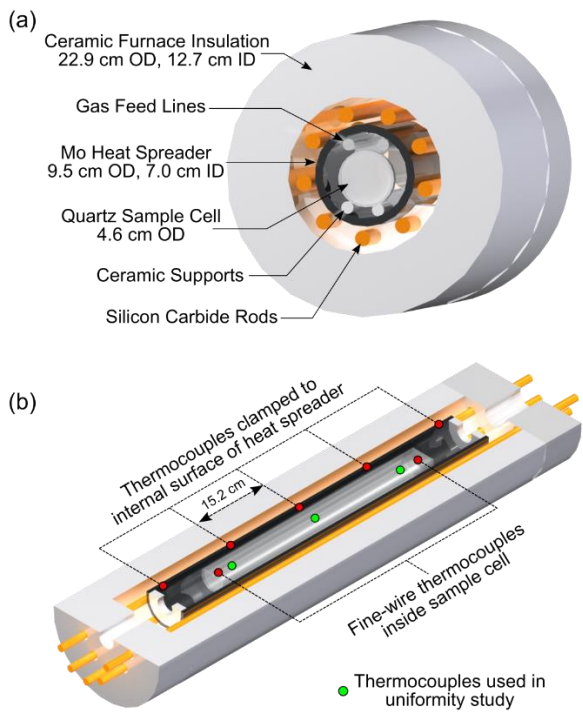

**Figure 2**: Detailed view of the ceramic furnace, heat spreader, and sample cell. Panel (a) shows a cross section of the furnace along with critical dimensions. Panel (b) shows the approximate location of the seven K-type thermocouples used to monitor the temperature distribution of the sample gas.

As mentioned above, the quartz sample cell is maintained on a separate gas line from the bath gas in order to maintain a slight pressure differential between the sample cell and the bath gas. The sample gas pressure is controlled using an electronic pressure regulator (Equilibar EPR3000) that also drives a spring-biased differential pressure regulator (TESCOM 26-2000) that controls the bath gas pressure and is set to maintain a 0.5 to 1.5 bar pressure differential over the sample cell pressure. As mentioned above, by controlling the pressure in this way, the quartz sample cell experiences only a slight pressure differential relative to the bath gas surrounding it, regardless of the absolute pressures in the gas cell.

To operate the experiment, the furnace temperature is first stabilized at a desired set point. The sample cell pressure is then slowly increased while monitoring the bath gas pressure to ensure that a slight positive pressure differential is maintained at all times. Once the desired pressure is reached, the temperature distribution must stabilize again after the addition of the cool gas. Light from the spectrometer is then passed axially through the gas cell to record the sample gas spectrum at the conditions in the sample cell.

## 4. Performance

### a. Temperature uniformity

As discussed above, accurate spectroscopy at high pressure and temperature requires the sample gas to be maintained at uniform, precisely known temperatures for the duration of the experiment. To assess the temperature uniformity of the gas cell, we installed a separate quartz cell with three internal K-type thermocouples placed along the center axis of the cell (see Figure 2). We then operated the gas cell over a matrix of pressure and temperature conditions to cover the entire range of expected experimental conditions. At each condition, we stabilized the gas cell for approximately one hour to allow the system to reach steady state. We recorded uniformity measurements at three temperatures: 500 K, 750 K, and 1000 K. At each temperature, we stabilized the gas cell at pressures of 1, 2, 5, 10, 15, 20, 30, 40, and 50 bar. At 500 K, we also measured the uniformity at 100 bar. All reported temperature readings are corrected for radiation effects (see Appendix).

We fit the measured gas temperatures with a second-order polynomial to extrapolate the measured temperatures to the edge of the quartz cell. We then evaluate the temperature uniformity from the polynomial fit as the maximum temperature deviation from the mean. We repeat this process for every measured temperature and pressure condition, and interpolate the resulting matrix of measured temperature uniformities using a bilinear algorithm to give a smooth estimate of the uniformity over the entire matrix of pressures and temperatures. Figure 3 shows the measured temperature uniformity of the gas cell as a function of the set temperature and pressure. Interestingly, Figure 3 shows that the temperature uniformity of the sample gas is a strong function of pressure, an effect that has not been measured in studies of prior gas cells. Temperature variations increase with pressure at lower pressures (<15 bar), and then decrease with increasing pressure after the variation reaches a maximum. A maximum temperature deviation of ~46 K occurs at 16 bar and 1000 K, which represents a ~4.5% deviation from the mean sample gas temperature.

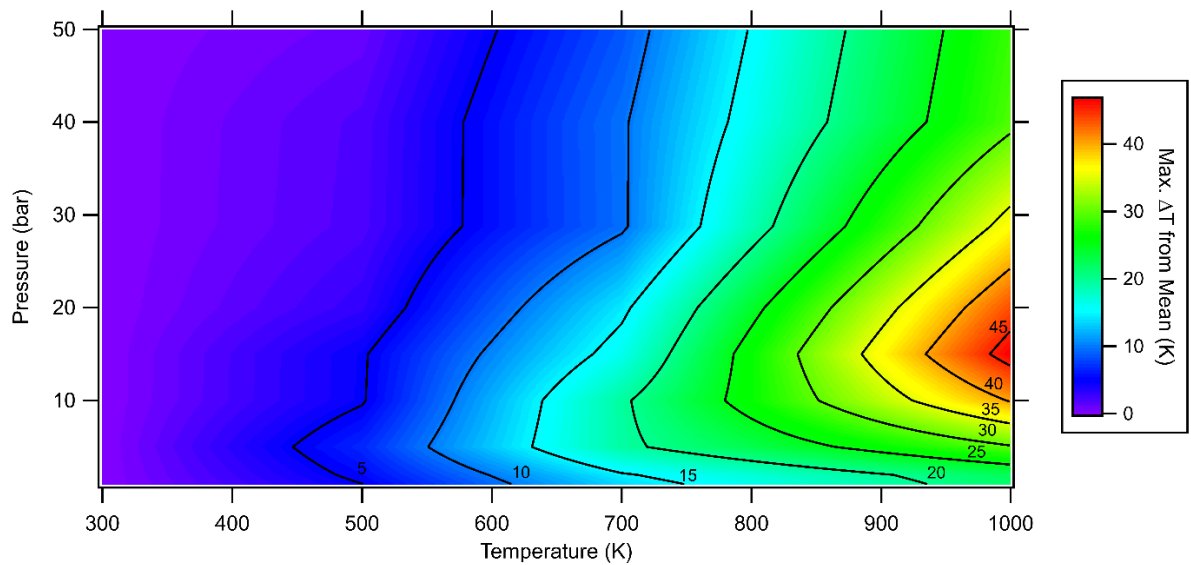

**Figure 3**: Visualization of the temperature uniformity (defined as the maximum temperature deviation from the mean) across a matrix of pressures and temperatures up to 50 bar and 1000 K. Uniformity values were measured at temperatures of 500 K, 750 K, and 1000 K and pressures of 1, 2, 5, 10, 15, 20, 30, 40, 50 bar. The resulting data is interpolated using a bilinear algorithm to give a smooth estimate of the temperature uniformity across the entire matrix of conditions.

The uniformity measurements also reveal a pressure-dependent drop in the sample gas temperature relative to the measured heat spreader temperature. Although the sample gas temperature assumes the same relative temperature distribution as the heat spreader, the peak temperature of the sample gas is significantly lower than that of the heat spreader. Figure 4 plots this temperature drop, which is linear with the gas cell pressure. Notably, Figure 4 shows that the measured heat spreader temperatures cannot be used to directly infer the sample gas temperature. Instead, the true peak gas temperature is determined using the linear fit parameters shown in Figure 4, while the edge temperatures of the sample cell are continuously monitored using the *in-situ* fine-wire thermocouples.

Finally, it is interesting to compare the temperature uniformity of the present cell design to the three most-recent gas cells with published uniformity data [19–21]. In Ref. [19], the authors assess temperature uniformity using internal, fine wire thermocouples and report ±1-2 K temperature variations across the 3 cm absorbing path length at 473 K and 1000 K. Both Refs. [20] and [21] infer gas temperature using thermocouples mounted on the exterior walls of the pressure vessel. Using this method, Ref. [20] reports 19 K deviations from the mean temperature along the 21.3 cm absorbing path length at 802 K, and Ref. [21] reports variations between 7 K and 10 K along the 9.4 cm path length for temperatures between 542 K and 1227 K. Notably, none of the prior studies report a pressure dependence to the measured temperature uniformity.

To account for the differing path lengths, we compare the different gas cells using a path length-normalized temperature uniformity (the maximum temperature deviation from the mean per unit path length). Figure 5 shows a comparison of the path length-normalized uniformity for the four gas cells as a function of temperature at ambient pressure. The present gas cell design demonstrates the highest uniformity (lowest temperature deviation per unit path length) for temperatures less than ~900 K. Above 900 K, the design of Christiansen *et al.* [19] is most uniform.

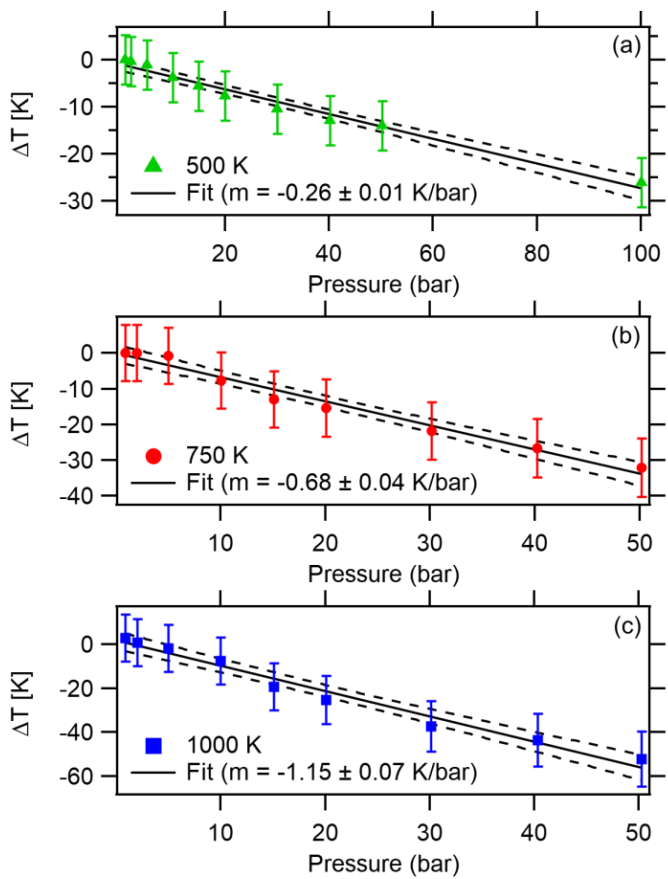

**Figure 4**: Shows the drop in the peak sample gas temperature relative to the heat spreader temperature as a function of pressure. Data were measured at temperatures of 500 K (a), 750 K (b), and 1000 K (c). Dashed lines show $1\sigma$ confidence intervals around the linear fit (solid line).

### b. Pressure stability

The pressure stability of the gas cell also influences the accuracy of spectroscopic measurements. In this cell design, the sample cell is maintained at a slightly lower pressure than the bath gas for the duration of the experiment. As such, leaks in the sample gas line can introduce a small amount of the non-absorbing bath gas into the sample cell. These leaks manifest as a slow increase in the sample gas pressure while the gas cell is maintained at static conditions. As the gas cell is often run continuously for several hours, it is important to assess the effect of minor leaks on the purity of the sample gas.

To assess the pressure stability of the gas cell, we measure the leak rate into the sample cell over a range of pressures and temperatures up to 1000 K. As the pressure is varied, the differential pressure between the bath gas and the sample cell (which drives the leak rate into the sample cell) varies, typically between 0.5 and 1.4 bar. For all the conditions tested, we measure an average leak rate of 0.16 mBar/min. With this leak rate, the mole fraction of a pure absorbing gas at 10 bar would be diluted by <0.3% over the duration of a three hour experiment.

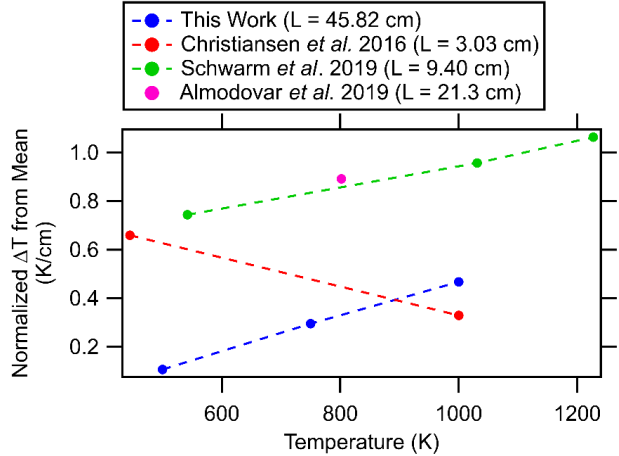

**Figure 5:** Compares the path length-normalized temperature uniformity of the present gas cell to three recently-published gas cell designs [19-21]. The present gas cell design exhibits the best uniformity for temperatures less than 900 K, while the design of Christiansen *et al*. [19] is most uniform above 900 K.

*c. Maximum achievable temperature and pressure conditions*

The maximum achievable temperature and pressure conditions within the sample gas are limited by the rating of the custom pressure vessel (100 bar at 300 °C skin temperature) and the maximum furnace power (16.9 kW). To estimate the peak pressure and temperature, we track the pressure vessel temperature and the current drawn by the furnace while operating the gas cell across a range of pressure and temperature conditions. Figure 6 shows the measured pressure vessel skin temperature and furnace current, which are both approximately linear with pressure. Extrapolation of the measured data shows that the gas cell is expected to reach the 100 bar pressure limit for sample gas temperatures up to 750 K. For a sample gas temperature of 1000 K, we expect the gas cell to be capable of ~80 bar before both the pressure vessel skin temperature and furnace power reach their limiting values.

The estimates above give the maximum sample gas temperature and pressure based on the physical constraints of the gas cell. In reality, other factors can limit the peak pressure and temperature conditions in which accurate reference spectroscopy can be performed. For example, in earlier iterations of this gas cell design, we found that convection in the high-temperature, high-pressure bath gas at either end of the sample cell caused significant beam steering though the gas cell, which in turn significantly reduced the spectrometer signal-to-noise ratio. In the present design, quartz tubes were placed along the center axis of the gas cell to reduce the exposure of the laser beam to convective currents within the bath gas.

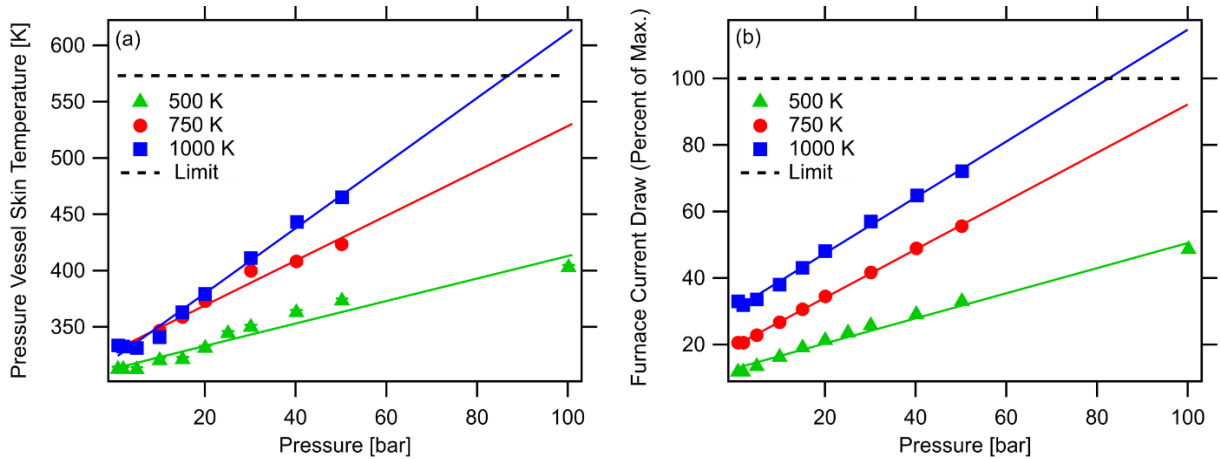

**Figure 6**: Panel (a) shows the skin temperature of the pressure vessel as a function of pressure for furnace set points of 500 K, 750 K, and 1000 K. Panel (b) shows the current draw (as a percentage of the maximum) as a function of pressure for the three furnace set points. Both panels indicate that the gas cell is capable of ~80 bar at 1000 K before reaching a hardware limit.

## 5. Dual frequency comb spectroscopy of the $CO_2$ $3\nu_3$ band

To further test and demonstrate the performance of the gas cell, we measure the spectrum of $CO_2$ between 6800 and 7000 cm$^{-1}$ using a dual frequency comb spectrometer. The dual-comb spectrometer used in these measurements [25–27] is based on fiber mode-locked combs with a spectral point spacing of 0.0066 cm$^{-1}$ and a broadened spectrum that spans ~5500-8000 cm$^{-1}$. Both self-referenced frequency combs are stabilized to a known cw reference laser with locks that are referenced to a GPS disciplined oscillator (Jackson Labs Fury). Through the unique combination of the long path length gas cell and the broadband, high-resolution spectrometer, we measure spectra spanning more than two orders of magnitude in pressure at temperatures up to 1000 K.

Using the dual-comb spectrometer, we coherently average [28] spectra for the pure $CO_2$ sample (Airgas Instrument Grade >99.99% purity) for a period of 30-60 minutes to increase the signal to noise ratio. We normalize each measured spectrum using a reference spectrum measured while the gas cell is under vacuum as well as a second reference spectrum measured concurrently with the $CO_2$ spectrum. The latter reference spectrum is measured though the same optical components as the gas cell (quartz and sapphire windows), but passes only though a region of purged dry air outside of the gas cell. In practice, residual baseline variation (<5%) remains after this normalization process due to drift in the output intensity spectrum of the dual-comb spectrometer and the optical alignment. We remove this remaining baseline variation by fitting each spectrum with a single Chebyshev polynomial (4-20 coefficients). Finally, although all portions of the optical path are purged with dry air during the measurement, a small amount of residual water vapor persists in the gas cell. Background water vapor absorption is subtracted from the $CO_2$ spectra through a fit to the water lines with a Speed-Dependent Voigt profile and spectral parameters for Argon-broadened water [29].

Figure 7(a) shows representative $CO_2$ spectra collected at ~732 K over a range of pressures between 0.2 and 20 bar. Figures 7(b) and (c) highlight the region around the $3\nu 3$ (00031 ← 00001) bandhead. Panel (b) shows the bandhead region measured at 203.0 ± 0.5 mbar and an average gas temperature of 731 ± 10 K. The dashed line shows the predicted absorption calculated using the HITRAN 2016 database and a Voigt profile [30]. Although the agreement between the measured data and the HITRAN model is good, the remaining residuals are dominated by errors in the linewidth. This is likely due to the fact that HITRAN 2016 does not include the temperature dependence of self-broadening for $CO_2$ in this region (here we have assumed $n_{self} = 0.5$ for all lines). In fact, there are very few prior measurements of self-induced broadening and shift for this band [31–33], and temperature dependence has only been measured for eight lines in the R-branch [33].

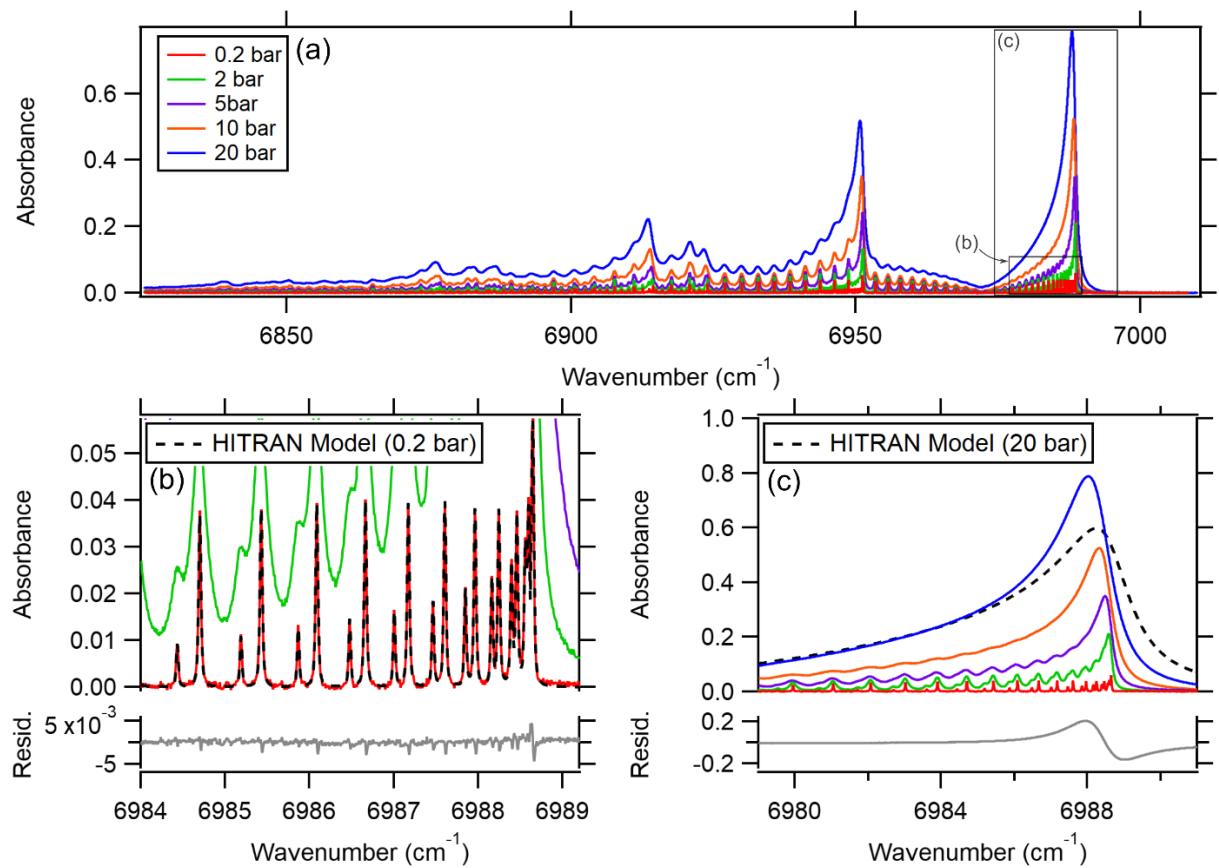

**Figure 7**: Panel (a) shows $CO_2$ spectra recorded in the new gas cell at a mean gas temperature of ~732 K using a dual frequency comb spectrometer with 0.0066 cm$^{-1}$ point spacing. Panel (b) shows a detail of the $3\nu_3$ bandhead region measured at 203.0 ± 0.5 mbar as well as a HITRAN-based model at the same conditions. Panel (c) shows the same bandhead region at ~20 bar. Here, there is poor agreement between the measured spectrum and the HITRAN model due to collisional effects (e.g. line mixing) that are not included in the HITRAN model.

Figure 7(c) shows the $3\nu_3$ bandhead region at high pressure. Here, we compare the bandhead measured at 19.8 ± 0.1 bar and an average gas temperature of 730 ± 10 K to a HITRAN 2016 model calculated at the same conditions [30]. As expected, the agreement between the measured spectrum and the HITRAN model is quite poor, with differences exceeding 20%. We attribute the large discrepancy to two effects. First, as mentioned above, the HITRAN model does not include the self-induced pressure shift as well as the temperature dependence of the shift and broadening. The omission of these parameters can lead to significant differences in the absorption profile when the model is extrapolated to very high pressures and temperatures. Secondly, it is well known that line mixing is a strong effect at high pressures, particularly in bandhead regions where absorption features are closely spaced [4,5]. In Figure 7(c), the HITRAN model strongly under predicts the measured spectrum at the bandhead while over estimating the absorption in the low-frequency side of the R-branch. This pattern has been observed in past studies of this band at elevated pressure [22,34–37], and is expected in a region experiencing strong line mixing more generally because line mixing results in a shift in absorption from weakly to strongly absorbing regions [4,5]. Figure 7(c) also shows that the HITRAN model over predicts absorption beyond the bandhead. While previous line mixing studies of pure $CO_2$ show that line mixing can replicate this effect relatively close to the bandhead, absorption far from the bandhead is due to the far wings of the R-branch lines [22,38]. In the far wings, absorption models must also consider the breakdown of the impact approximation (i.e. consider the finite duration of collisions) in order to accurately model absorption at high densities [15,22,39,40].

As a final example of the spectra measured using the new gas cell, Figure 8 shows the evolution of the $CO_2$ spectrum with temperature at ~14.8 bar. At 296 K, the spectrum is dominated by the $3\nu_3$ band, with only a weak contribution from a hot band near 6935 cm$^{-1}$ (01131 ← 01101). As temperature is increased, the prevalence of the $01131 \leftarrow 01101$ hot band increases dramatically, and absorption shifts further to lower frequencies as a larger number of nearby hot bands become activated. At a mean gas temperature of 964 K, bandheads for more than ten individual absorption bands are visible in the high pressure spectrum, some of which are indicated in Figure 8. Many of these hot bands overlap in the high-pressure, high-temperature spectrum, further complicating absorption models for these conditions. This ability to resolve numerous, overlapping bands in a single spectrum highlights a unique benefit of combining the high-pressure, high-temperature gas cell with the broadband dual frequency comb spectrometer.

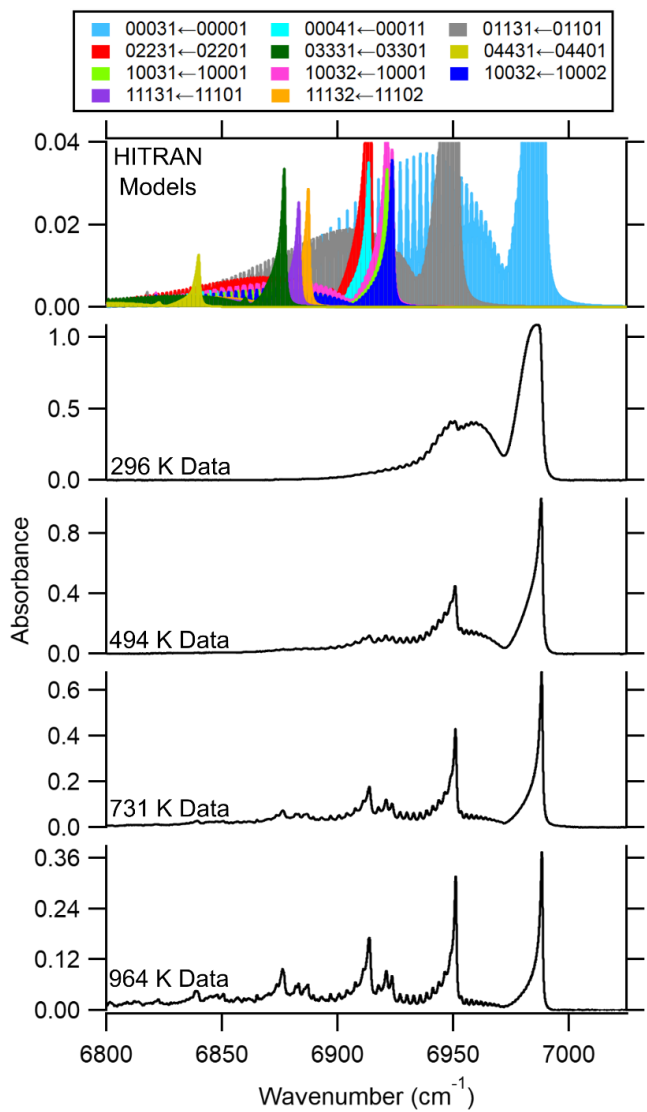

**Figure 8**: Shows the evolution of the measured $CO_2$ spectrum at 14.8 bar for temperatures between 296 K and 964 K. The top panel shows many of the hot bands (simulated at 964 K and 14.8 bar) that overlap in the high pressure, high temperature spectra. The different bands are labeled as $v'_1 v'_2 l'_2 v'_3 n' \leftarrow v''_1 v''_2 l''_2 v''_3 n''$.

## 6. Summary and conclusion

We present the design and characterization of a new gas cell for absorption spectroscopy at high pressure and temperature. The design is based on a carefully-controlled quartz sample cell housed inside of a custom furnace and pressure vessel that are used to generate the high-pressure and -temperature environment. We surround the glass sample cell with a molybdenum heat spreader that significantly increases the temperature uniformity across the 45.82 cm absorbing path. We validate the gas cell by measuring the temperature uniformity of the sample gas using *in-situ* thermocouples across a matrix of pressure and temperature conditions up to 50 bar at 1000 K. The measurements demonstrate that variations in the sample gas temperature remain within 4.5% of the mean temperature across the entire range of pressure and temperature conditions. Additionally, the measurements identify pressure-dependent effects in the sample gas temperature that must be considered in order to accurately interpret measured spectra. Taken as a whole, our characterization of the new gas cell demonstrates that the new cell design is capable of generating a precisely known and highly uniform, high-pressure and -temperature environment over an absorbing path length significantly longer than existing gas cells.

Uniquely, we combine the new gas cell with a dual frequency comb absorption spectrometer to record broadband, high resolution spectra of $CO_2$ between 6800 and 7000 cm$^{-1}$. The measurements identify deviations from the HITRAN 2016 absorption model at low pressures due to the lack of self-induced collisional parameters and their temperature dependence for the measured $CO_2$ bands in the HITRAN database. Further, the measurements highlight additional collisional effects (line mixing and the breakdown of the impact approximation) that cause significant deviations at high pressure from a model based only on a Voigt line shape profile. These results motivate the need for further laboratory spectroscopy at high pressure and temperature to improve absorption databases and models for applications in combustion and planetary science.

## 7. Acknowledgements

This work was supported by the Air Force Office of Scientific Research through grant number FA9550-17-1-0224. R. Cole acknowledges support from NASA Headquarters under the NASA Earth and Space Science Fellowship Program (PLANET17F-0120, 18-PLANET18R-0018, 19-PLANET19R-0008). The authors would like to thank Mr. Arun Kumar, Mr. Daniel Llorca and Mr. George Carter for their help in constructing the gas cell.

## 8. Appendix: Thermocouple Radiation Correction and Uncertainty Analysis

We apply a correction to our measured thermocouple temperatures to account for radiation effects and to ensure that each measured temperature reflects the true gas temperature. However, it is important to note that, because all of the thermocouples are located inside of the molybdenum heat spreader, the magnitude of radiation correction is relatively small in all cases because there is no line-of-sight between the thermocouples and a bright surface (e.g. a heating element).

We account for radiation in our measured temperatures by equating the net convective heat transfer between the thermocouple bead and the gas to the radiation exchange between the thermocouple bead and the inside surface of the molybdenum heat spreader [41,42]. To account for the temperature distribution of the heat spreader, we discretize the heat spreader into five segments, as shown in Figure 9. We fix the temperature of each segment at the temperature measured by the five thermocouples clamped to the inside surface of the heat spreader (see Figure 2). With this framework, the radiation-corrected gas temperature is given by

$$T_{gas} = \frac{1}{h}\left[\sum_{i=1}^{5} \epsilon_{TC}\sigma F_{TC,i}(T_{TC}^4 - T_i^4)\right] + T_{TC} \tag{1}$$

where $h$ is the convective heat transfer coefficient, $\epsilon_{TC}$ is the emissivity of the thermocouple bead, $\sigma$ is the Stefan-Boltzmann constant, $F_{TC,i}$ is the view factor between the thermocouple bead and segment $i$ of the heat spreader, $T_{TC}$ is the measured thermocouple temperature, and $T_i$ is the measured temperature of segment $i$ of the heat spreader. For this analysis, we assume that the heat spreader is a sealed enclosure, and therefore neglect heat transfer through the small holes (~2.5 cm) at either end of the heat spreader. Further, this model does not include the effect of the quartz cell surrounding the thermocouples. Including the quartz cell in the heat transfer calculation (modeled as a radiation shield [41,42]) negligibly affects the magnitude of the radiation correction.

To calculate the gas temperature, we adopt a value for the thermocouple emissivity of 0.1 [43], and calculate the view factors between the thermocouple and each segment of the heat spreader using Engineering Equation Solver [44,45]. Figure 9 gives values for each view factor. We estimate the convection coefficient as $h = 20 \pm 5 \frac{W}{m^2 K}$ using the scaling relations for a horizontal cylindrical enclosure given in [46]. To calculate the uncertainty in the magnitude of the radiation correction, we calculate the gas temperature using the limiting values of $h = 25 \frac{W}{m^2 K}$ and $h = 15 \frac{W}{m^2 K}$, and define the uncertainty as the spread in the two resulting temperatures. We add this uncertainty in quadrature with the standard uncertainty for a K-type thermocouple (0.75% of reading) to assess the total uncertainty in the radiation-corrected gas temperature. As mentioned above, the magnitude of the radiation correction is relatively small. For example, at a set point temperature of 1000 K, the radiation correction results in a change in the thermocouple-measured gas temperature of ~3 K at ambient pressure, and ~25 K at 50 bar.

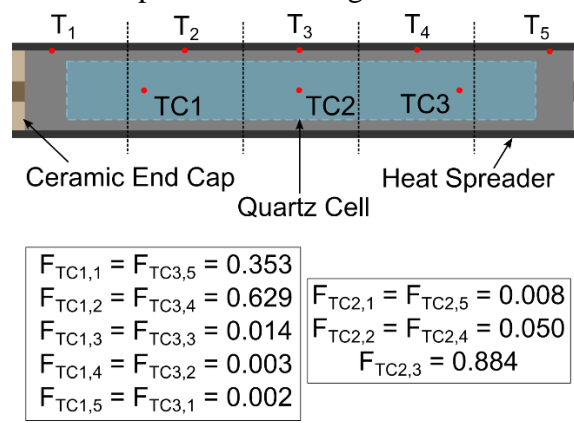

**Figure 9**: Schematic of the heat spreader and quartz cell used for the thermocouple radiation correction. Red dots show the locations of thermocouples measuring the sample gas and the heat spreader temperature. View factors are given between the three sample gas thermocouples and the five segments of the heat spreader.